\documentclass{kapproc}

\usepackage{amsmath, amsfonts}

\let\footnote\savefootnote
\let\footnotetext\savefootnotetext
\let\footnoterule\savefootnoterule

\normallatexbib

\begin{document}

\articletitle
[Sine-Gordon Field with Boundary Oscillator]
{Coupling the Sine-Gordon Field \\
Theory to a Mechanical System \\
at the Boundary}

\author{P.Baseilhac \footnote{pascal@yukawa.kyoto-u.ac.jp}}

\affil{
Yukawa Institute for Theoretical Physics, Kyoto University, 606-8502
Kyoto, Japan.}

\author{G.W.Delius \footnote{gwd2@york.ac.uk; delivered talk}
and A.George \footnote{ag160@york.ac.uk; author of paper}}

\affil{
Department of Mathematics, University of York,
Heslington, York, YO10 5DD, UK
\footnote{http://www.york.ac.uk/mathematics/physics/}}

\begin{abstract}
We describe an integrable system consisting of the sine-Gordon
field, restricted to the half line, and coupled to a non-linear
oscillator at the boundary.  By extension
of the coupling constant to imaginary values we also outline
the equivalent system for the sinh-Gordon field.  We show how
Sklyanin's formalism can be applied to situations with dynamic boundary
conditions, and illustrate the method with the derivation of our example
system.

\end{abstract}

\section{Introduction}

Recent work on integrable quantum fields in 1+1 dimensions has introduced a
class of theories where the domain of the field is bounded.  The
boundaries restrict
the field either to the half line or an interval and impose fixed
boundary conditions upon it (see, for example, \cite{delref13, delref23,
delref24}).  In the present work we introduce a new class of integrable
theories where
the field is coupled to a classical or quantum
mechanical system at the boundary of its domain. Effectively
we are allowing new degrees of freedom to exist on the
boundary and so the field has dynamic boundary conditions.

In section two we describe one example of such a system, where the
sine-Gordon field is restricted to the half line
bounded by a non-linear oscillator. Then by extension of the
coupling constant into imaginary values we briefly consider
the equivalent system for the sinh-Gordon field.

Section three contains the derivation of our example system, including
details on how Sklyanin's technique has applied to a system with
dynamic boundary K matrices. Section four comprises a discussion on
this class of theory.

Much of the content of this paper follows the work of Baseilhac and
Delius \cite{gustav}.  Due to restrictions on space we have
omitted from the present work the discussion on dynamic solutions
to the quantum
reflection equation that appeared in \cite[section 5]{gustav}.

\section{The Sine-Gordon Field Coupled to Oscillator at Boundary}

In this section we describe an integrable systems consisting of
a sine-Gordon field restricted to the half line and coupled to an
oscillator at the boundary.

The sine-Gordon model is a 1+1 dimensional theory describing a
relativistic, self-interacting, massive, bosonic field.
We introduce a boundary at the origin by restricting the field to the half
line between $x = -\infty$ and $x = 0$. The Hamiltonian for such a field is
\begin{equation}
\label{hsg}
H_{\text{sG}} =
\int_{-\infty}^0 \left(
    \frac{1}{2} \pi^2 +
    \frac{1}{2} \left( \partial_x \phi \right)^2 -
    \frac{m^2} {\beta^2}
    \left( \cos \beta \phi - 1 \right)
\right) dx,
\end{equation}
where $\phi(x,t)$ is the sine-Gordon field and $\pi(x,t)$ is the
conjugate momentum, with the usual Poisson brackets $\{ \pi(x) , \phi(y)
\} = \delta (x-y)$. Both $\beta$, the sine-Gordon coupling
constant, and $m$, the mass scale parameter, are real.

We now couple
the field at the boundary to a classical system with the Hamiltonian
\begin{equation}
\label{hosc}
H_{\text{osc}} =
-\frac{2m}{\beta^2} \left(
    e^{-i \beta \phi (0) / 2}
    \cos \hat{p} +
    e^{i \beta \phi (0) / 2}
    \cos \hat{q}
\right),
\end{equation}
where $\phi (0)$ is the field at $x = 0$, and $\hat{q}$ and $\hat{p}$
are related to the position, $q$,
and momentum, $p$, variables of the oscillator as
\begin{equation}
\label{pq}
\hat{p} = \frac{\beta}{\sqrt{2Mm}} p,
~~~~ \text{and} ~~~~
\hat{q} = \frac{\beta \sqrt{Mm}}{2 \sqrt{2}} q,
\end{equation}
where $M$ sets the mass of the oscillator. This becomes obvious after
expansion of \eqref{hosc} for small $p$ and $q$.
Position and momentum variables have the usual Poisson bracket $ \{ p,q
\} =1$ and our rescaled variables have the relation $ \{ \hat{p} ,
\hat{q} \} = \beta^2 / 4$.

The total Hamiltonian of the system can be expressed as the sum of
the oscillator and bulk field Hamiltonians
\begin{equation}
\label{ham1}
H = H_{\text{sG}} + H_{\text{osc}}.
\end{equation}

We are now in a position to find Hamilton's equations for the bulk
field, they are
\begin{align}
\label{heph}
\frac{d}{dt} \phi (x) = &
\{ H, \phi (x) \} = \pi (x), \\
\label{hepi}
\frac{d}{dt} \pi (x) = &
\{ H, \pi (x) \} = \partial_x^2 \phi (x) - \frac{m^2}{\beta}
\sin \beta \phi (x) \\
& - \delta (x) \left( \partial_x \phi (0) + \frac{im}{\beta} \left( e^{-i
\beta \phi (0) / 2} \cos \hat{p} - e^{i \beta \phi (0) / 2} \cos \hat{q}
\right) \right).
\nonumber
\end{align}
The two terms proportional to $\delta (x)$ in \eqref{hepi} come from two
different sources.  The $\partial_x \phi (0)$ term comes from the
boundary contribution due to the partial integration of the sine-Gordon
field. The other terms are due to the boundary oscillator containing
$\phi (0)$. We require, for a physical solution, that $\pi (x)$ is
continuous, i.e. the $\delta (x)$ term must be zero. This gives us the
boundary condition
\begin{equation}
\label{bound}
\partial_x \phi (0) =
- \frac{im}{\beta} \left( e^{-i \beta \phi (0) /2} \cos \hat{p} - e^{i
\beta \phi (0) /2} \cos \hat{q} \right).
\end{equation}

By substituting \eqref{heph} into \eqref{hepi} and applying the
boundary condition \eqref{bound} we recover the usual sine-Gordon
equation of motion
\begin{equation}
\label{eqm}
\partial_t^2 \phi - \partial_x^2 \phi = -\frac{m^2}{\beta} \sin \beta
\phi.
\end{equation}

We can find Hamilton's equations for the boundary oscillator in a
similar way
\begin{equation}
\label{hep}
\frac{d}{dt} \hat{p} =
\{ H, \hat{p} \} =
- \frac{m}{2} e^{i \beta \phi (0) /2} \sin \hat{q},
\end{equation}
\begin{equation}
\label{heq}
\frac{d}{dt} \hat{q} =
\{ H, \hat{q} \} =
\frac{m}{2} e^{-i \beta \phi (0) /2} \sin \hat{p}.
\end{equation}
By combining these two equations we are able to find the equation of
motion of the oscillator
\begin{equation}
\label{emos}
\left( \ddot{\hat{q}} +
\frac{i \beta}{2} \dot{\phi} (0) \dot{\hat{q}}
\right)
\mathcal{Z}
\left(
\frac{2}{m} e^{i \beta \phi (0)/2} \dot{\hat{q}}
\right) =
-\frac{m^2}{4} \sin \hat{q},
\end{equation}
where dots denote derivatives with respect to time and the function
$\mathcal{Z}$ is defined as
\begin{equation}
\label{defz}
\mathcal{Z} (x) =
\frac{d}{dx} \sin^{-1} (x) = \frac{1}{\sqrt{1-x^2}}.
\end{equation}
We note that the oscillator's Hamiltonian, and thus the boundary
condition for the field, are complex.  This generally leads to complex
solutions, both in the bulk and for boundary states.  This is not
without precedent, for example soliton solutions in affine Toda theories
are complex, however it has been shown that the energies of these
solutions are nevertheless real \cite{delref15, olive}.
We hope that the energies of solutions to our system will also be real.

\bigskip

We note that our boundary condition \eqref{bound} is similar in form to
the previously known integrable boundary conditions \cite{delref23,
delref13, delref19}
\begin{equation}
\label{pcbc}
\partial_x \phi (0) =
- \frac{im}{\beta} \left( \eta_0 e^{-i \beta \phi (0)/2} -
\eta_1 e^{i \beta \phi (0)/2} \right),
\end{equation}
where $\eta_0 ~ \text{and} ~ \eta_1$ are parameters.  In the classical
case our boundary condition \eqref{bound} reduces to equation
\eqref{pcbc} with $\eta_0  = \eta_1 = 1$ when $p = q = 0$.  This will
never occur in the quantum case as the position and momentum of the
system can never be simultaneously known.

\bigskip

We also note that by requiring the coupling constant to be purely
imaginary, $\beta = i \hat{\beta}$, we can obtain the equations for
the equivalent integrable system for the sinh-Gordon equation.  For
instance the Hamiltonian becomes
\begin{align}
H_{ShG} &  =
\int_{-\infty}^0 \left( \frac{1}{2} \pi^2 +
\frac{1}{2} \left( \partial_x \phi \right)^2 +
\frac{m^2}{\hat{\beta}^2}
\left( \cosh \hat{\beta} \phi - 1 \right)
\right) dx \\ \nonumber
\label{shg}
+ & \frac{2m}{\hat{\beta}^2} \left(
e^{-\hat{\beta} \phi (0) /2}
\cosh \left( \frac{\hat{\beta}}{\sqrt{2Mm}} p \right) +
e^{\hat{\beta} \phi (0) /2}
\cosh \left( \frac{ \hat{\beta} \sqrt{Mm}}{2 \sqrt{2}} q \right)
\right).
\end{align}
We note that the Hamiltonian is real, which leads to real solutions.

\bigskip

\section{Derivation of Sine-Gordon Example}

In this section we describe Sklyanin's formalism \cite{delref23} and how
we have applied this technique to dynamic boundary conditions.  We
illustrate our method with the derivation of the example system
presented in section two.

As required by Sklyanin's formalism there exist two matrix valued
functions $a_x ( \theta , x)$ and $a_t (\theta , x)$ which depend upon
the sine-Gordon field, its conjugate momentum, and on the spectral
parameter $ \theta \in \mathbb{C}$ such that the classical equation
of motion for the field \eqref{eqm} can be written as the Lax pair
equation
\begin{equation}
\label{laxp}
[ \partial_x - a_x ( \theta , x), \partial_t - a_t ( \theta, x) ] = 0
~~~~ \forall ~ \theta.
\end{equation}
These are
\begin{align}
a_x ( \theta , x) =
\frac{\beta}{4i} \frac{ \partial \phi}{ \partial t} \sigma_3
& +
\frac{m}{4i} \left( e^{\theta} + e^{-\theta} \right) \sin \left(
\frac{\beta \phi}{2} \right) \sigma_1 \nonumber \\
\label{ax}
& +
\frac{m}{4i} \left( e^{\theta} - e^{-\theta} \right) \cos \left(
\frac{\beta \phi}{2} \right) \sigma_2 ,
\end{align}
\begin{align}
a_t ( \theta , x) =
\frac{\beta}{4i} \frac{ \partial \phi}{ \partial x} \sigma_3
& +
\frac{m}{4i} \left( e^{\theta} - e^{-\theta} \right) \sin \left(
\frac{\beta \phi}{2} \right) \sigma_1 \nonumber \\
\label{at}
& +
\frac{m}{4i} \left( e^{\theta} + e^{-\theta} \right) \cos \left(
\frac{\beta \phi}{2} \right) \sigma_2 ,
\end{align}
where $\sigma_k$ are the Pauli matrices,
$\sigma_1 = \left( \begin{smallmatrix}
            0 & 1 \\
            1 & 0
        \end{smallmatrix} \right),
~\sigma_2 = \left( \begin{smallmatrix}
            0 & -i \\
            i & 0
        \end{smallmatrix} \right),
~\sigma_3 = \left( \begin{smallmatrix}
            1 & 0 \\
            0 & -1
        \end{smallmatrix} \right)$.

Equation \eqref{laxp} is the compatability condition for the
overdetermined set of equations
\begin{align}
\label{eqsys}
\frac{ \partial T}{ \partial x_{+}} & = a_x ( \theta , x_{+} ) T, \\
\frac{ \partial T}{ \partial t} & = a_t (\theta , x_{+} ) T -
T a_t ( \theta , x_{-} ).
\nonumber
\end{align}
The transition matrix $ T \equiv T ( x_{+}, x_{-}, \theta )$ is defined
as the solution to \eqref{eqsys} with initial condition $T ( x_{-},
x_{-}, \theta) = I$. $T$ can be expressed as a path ordered exponential
\begin{equation}
\label{path}
T ( x_{+}, x_{-}, \theta ) = \mathcal{P} \exp \left( \int_{x_-}^{x_+}
a_x ( \theta, x) dx \right),
\end{equation}
such that the operators are ordered with those at points nearest $x_+$
furthest to the left. This gives the matrix the inversion property
\begin{equation}
\label{invt}
T^{-1} ( x_+ , x_- , \theta ) = T ( x_- , x_+ , \theta ).
\end{equation}

The Poisson brackets for the matrices $a_x ( \theta, x)$
can be written in the form
\begin{align}
\label{pba}
\{ \stackrel{1}{a}_x ( \theta , x ) ,
\stackrel{2}{a}_x ( \theta' , y )
\} = & \\ \nonumber
\delta (x-y)
[ r& ( \theta - \theta' ) ,
\stackrel{1}{a}_x ( \theta , x ) +
\stackrel{2}{a}_x ( \theta' , y ) ] ,
\end{align}
which leads to Poisson brackets for the transition matrix
\begin{align}
\label{pbt}
\{ \stackrel{1}{T} ( x_+ , x_- , \theta ),
\stackrel{2}{T} ( x_+ , x_- , \theta' ) \}
= & \\ \nonumber
[ r(\theta - \theta' ), &
\stackrel{1}{T} ( x_+ , x_- , \theta)
\stackrel{2}{T} ( x_+ , x_- , \theta')],
\end{align}
where $ \stackrel{1}{A} = A \otimes I$ and $ \stackrel{2}{A} = I \otimes
A$.  This gives us the r-matrix, which is independent of both the field
and its conjugate momentum.

For the derivation of the sine-Gordon example we redefine the Poisson
brackets such that the range of the integral extends only over the space
occupied by the field. So at fixed time the Poisson brackets for any
observable $\mathcal{O}_j$ is defined as
\begin{align}
\{ \mathcal{O}_1 , \mathcal{O}_2 \} =
\int_{x_-}^{x_+} & \left(
\frac{\partial \mathcal{O}_1}{\partial \pi (x)}
\frac{\partial \mathcal{O}_2}{\partial \phi (x)} -
\frac{\partial \mathcal{O}_1}{\partial \phi (x)}
\frac{\partial \mathcal{O}_2}{\partial \pi (x)}
\right) dx
\nonumber \\
\label{defpb}
+ &
\frac{\partial \mathcal{O}_1}{\partial p}
\frac{\partial \mathcal{O}_2}{\partial q} -
\frac{\partial \mathcal{O}_1}{\partial q}
\frac{\partial \mathcal{O}_2}{\partial p}.
\end{align}
By calculating these Poisson brackets for $a_x (\theta, x)$ we can find
the r-matrix for the sine-Gordon system,
\begin{equation}
\label{rsg}
r_{\text{sg}} ( \theta ) =
\frac{\beta^2 \cosh ( \theta )}{16 \sinh ( \theta )}
( I \otimes I - \sigma_3 \otimes \sigma_3 ) -
\frac{ \beta^2 }{16 \sinh ( \theta )}
( \sigma_1 \otimes \sigma_1 + \sigma_2 \otimes \sigma_2 )
\end{equation}

We now introduce two matrix valued
functions of $ \theta$, $K_{\pm} ( \theta )$, which are independent of the
field and describe the nature of the boundaries at $x = x_{\pm}$.
It is at this point that we deviate from previous applications of
Sklyanin's method. We wish to describe a dynamic boundary so require
that its $K$ matrix be dependent upon the position and momentum
variables of the boundary system.
We further require that the $K$ matrices
satisfy the classical reflection equation
\begin{align}
\label{cre}
\{ \stackrel{1}{K}_{\pm} ( \theta),
\stackrel{2}{K}_{\pm} ( \theta' ) \} & =
[ r ( \theta - \theta' ),
\stackrel{1}{K}_{\pm} ( \theta)
\stackrel{2}{K}_{\pm} ( \theta') ] \\
\nonumber
+ &
\stackrel{1}{K}_{\pm} ( \theta )
r ( \theta + \theta' )
\stackrel{2}{K}_{\pm} ( \theta' ) -
\stackrel{2}{K}_{\pm} ( \theta' )
r ( \theta + \theta' )
\stackrel{1}{K}_{\pm} ( \theta ).
\end{align}
As we want the two boundaries to be independent
of each other, we also require that the relation
\begin{equation}
\label{indb}
\{ \stackrel{1}{K}_{\pm} ( \theta ),
\stackrel{2}{K}_{\mp} ( \theta' ) \} = 0
\end{equation}
be satisfied.

As we are deriving a system on the half line
we choose to place $x_-$ at $-\infty$ and have $K_-$ take the
trivial solution to the classical reflection equation \eqref{cre}, $K_- (
\theta ) =
I$. We impose $\phi (x_- , t) = 0 , \pi (x_- , t)
= 0$ when $x_- = -\infty$.
We note the independence
condition \eqref{indb} is automatically satisfied due to our choice of
$K_-$. We will take \cite{delref26}
\begin{align}
\label{kpls}
& K_+ ( \theta ) = \\
\nonumber
& 2
\begin{pmatrix}
\cosh ( \tilde{p} + \tilde{q} ) e^{\theta} -
\cosh ( \tilde{p} - \tilde{q} ) e^{-\theta} &
2 \sinh^2 ( \theta ) - 2 \sinh^2 ( \tilde{p} ) \\
2 \sinh^2 ( \tilde{q} ) - 2 \sinh^2 ( \theta ) &
\cosh ( \tilde{p} - \tilde{q} ) e^{\theta} -
\cosh ( \tilde{p} + \tilde{q} ) e^{-\theta}
\end{pmatrix}
,
\end{align}
where $ \{ \tilde{p} , \tilde{q} \} = \beta^2 / 8$.
This is a solution to \eqref{cre} with $ r ( \theta ) =
r_{\text{sg}} ( \theta )$ defined in \eqref{rsg}. We will position this
boundary at the origin, $x_+ = 0$.

Following Sklyanin's technique we now define the generating function to be
\begin{equation}
\label{gf}
\tau ( \theta ) = \text{tr}
\left( K_+ ( \theta ) T ( x_+ , x_- , \theta ) K_- ( \theta ) T^{-1} ( x_+ ,
x_- , -\theta ) \right).
\end{equation}
These generating functions are in involution, i.e. $ \{ \tau ( \theta ),
\tau ( \theta' ) \} = 0$,
and can be expanded around any
singularities in $T ( x_+ , x_- , \theta )$.
Such an expansion will produce an infinite number of
conserved quantities $I_n$, that are also in involution with each other.
One of these quantities will be identifiable
as the Hamiltonian of the system, implying that all $I_n$ are time
conserved.

Applying the inversion relation \eqref{invt} and the fact that we chose
$K_- = I$ to \eqref{gf} gives us an expression for the
generating function for the sine-Gordon example
\begin{equation}
\label{sggf}
\tau_{\text{sg}} ( \theta ) = \text{tr}
\left( K_+ ( \theta ) T ( 0, -\infty , \theta ) T ( -\infty , 0 ,
-\theta ) \right).
\end{equation}
We note from \eqref{ax} that the matrix $a_x ( \theta , x )$ has
singularities at $ | \theta | = \pm \infty $. Thus the transition matrix
$T$ also has two singularities. Performing a Laurent
expansion on $\tau ( \theta )$ around either of these points gives a
Laurent series, the coefficients of which give us an infinite number of
conserved quantities $I_n$ \cite{delref10},
\begin{equation}
\label{coeffs}
\ln \tau ( \theta ) - 2 \theta =
\sum_{n=-1}^{\infty}{I_n e^{-n \theta}}.
\end{equation}
The first non-trivial quantity is
\begin{equation}
\label{i1}
I_1 = -\frac{i \beta^2}{2m} H + \text{const},
\end{equation}
where
\begin{align}
H = & \int_{-\infty}^0 \left( \frac{1}{2} \pi^2 +
\frac{1}{2} ( \partial_x \phi )^2 -
\frac{m^2}{\beta^2} ( \cos ( \beta \phi ) -1 )
\right) dx
\nonumber \\
\label{ham2}
& + \frac{2m}{\beta^2}
\left( e^{i \beta \phi(0)/2} \cosh ( \tilde{p} + \tilde{q} )
+ e^{-i \beta \phi (0) /2} \cosh ( \tilde{p} - \tilde{q} )
\right),
\end{align}
which, under the canonical transformation
\begin{equation}
\label{cannon}
\tilde{p} + \tilde{q} \to i \hat{q},
~~~
\tilde{p} - \tilde{q} \to - i \hat{p},
~~~
\phi (x) \to \phi (x) + 2 \pi / \beta,
\end{equation}
reproduces the Hamiltonian of our example sine-Gordon system presented
in section two (equations \eqref{hsg} and \eqref{hosc}). $\hat{p}$ and
$\hat{q}$ are defined in equation \eqref{pq}.
The next non-trivial quantity after the Hamiltonian gives the spin 3
charge, applying the same canonical transformation,
\begin{align}
\nonumber
I_3 =
\int_{-\infty}^0 &
\frac{ \beta^4 }{16 m^3}
\left(
    \pi^4 +
    6 \pi^2 ( \partial_x \phi )^2 +
    ( \partial_x \phi )^4
\right)
-
\frac{ \beta^2}{m^3}
\left(
    (\partial_x \pi )^2 +
    ( \partial_x^2 \phi )^2
\right)
\\
\nonumber
& -
\frac{\beta^2}{4m}
\left(
    \pi^2 +
    5( \partial_x \phi )^2
\right)
\cos ( \beta \phi ) +
\frac{m}{8}
( \cos (2 \beta \phi ) -1 )
 ~~ dx
\\
\nonumber
& +
e^{3i \beta \phi (0) /2}
\left(
    \frac{1}{2} \cos \hat{q} +
    \frac{1}{6} \cos^3 \hat{q}
\right)
\\
\nonumber
& -
e^{i \beta \phi (0) /2}
\left(
    \frac{3}{2} \cos \hat{p} -
    \frac{1}{2} \cos^2 \hat{q} \cos \hat{p} +
    \frac{\beta^2}{2m^2} \pi^2 \cos \hat{q}
\right)
\\
\nonumber
& +
e^{-3i \beta \phi (0) /2}
\left(
    \frac{1}{2} \cos \hat{p} +
    \frac{1}{6} \cos^3 \hat{p}
\right)
\\
\nonumber
& -
e^{-i \beta \phi (0) /2}
\left(
    \frac{3}{2} \cos \hat{q} -
    \frac{1}{2} \cos^2 \hat{p} \cos \hat{q} +
    \frac{\beta^2}{2m^2} \pi^2 \cos \hat{p}
\right)
\\
\label{bigun}
& -
\frac{2i \beta}{m} \sin \hat{q} \sin \hat{p}.
\end{align}
In principle, similar expressions can be obtained for any of the
infinite number of conserved charges of our system. The existence of
these higher order local integrals of motion shows that our system is
classically integrable.

\bigskip

\section{Discussion}

In this paper we have described in detail one example of an integrable
field theory coupled to a dynamic boundary system; this being the
sine-Gordon field, restricted to the half line, and coupled to a
non-linear oscillator at the boundary.

We have detailed how Sklyanin's formalism may be applied to systems
with dynamic $K$ matrices, and illustrated this with the derivation of
our example system.

Work on the sinh-Gordon and sine-Gordon systems presented here will
continue, treating the
systems both classically and in the quantum case.

\begin{acknowledgments}

We would like to thank
Evgueni Sklyanin, Vadim Kuznetsov and Peter Bowcock for helpful
discussions.
GWD is supported by an EPSRC advanced fellowship and AG by a PPARC
studentship.
PB's work was supported by Marie Curie fellowship HPMF-CT-1999-00094 and
is currently supported by a JSPS fellowship.
GWD thanks the organizers for a most enjoyable conference.

\end{acknowledgments}

\end{document}